\documentclass[conference]{IEEEtran}
\IEEEoverridecommandlockouts
\usepackage{cite}
\usepackage{amsmath,amssymb,amsfonts}
\usepackage{algorithmic}
\usepackage{graphicx}
\usepackage{textcomp}
\usepackage{xcolor}
\usepackage{multirow}
\usepackage{array}
\usepackage{dblfloatfix}
\usepackage{soul}
\usepackage{array, makecell}

\def\BibTeX{{\rm B\kern-.05em{\sc i\kern-.025em b}\kern-.08em
    T\kern-.1667em\lower.7ex\hbox{E}\kern-.125emX}}
\begin{document}

\title{Single-Shot Black-Box Adversarial Attacks Against Malware Detectors: A Causal Language Model Approach\\

\thanks{*: Corresponding author

Acknowledgments: This material is based upon work supported by the National Science Foundation (NSF) under Secure and Trustworthy Cyberspace (1936370), Cybersecurity Innovation for Cyberinfrastructure (1917117), and Cybersecurity Scholarship-for-Service (1921485) programs.}
}

\makeatletter
\newcommand{\linebreakand}{%
  \end{@IEEEauthorhalign}
  \hfill\mbox{}\par
  \mbox{}\hfill\begin{@IEEEauthorhalign}
}
\makeatother

\author{
\IEEEauthorblockN{James Lee Hu*}
\IEEEauthorblockA{\textit{Department of Management Information Systems} \\
\textit{University of Arizona}\\
Tucson, USA \\
jameshu@email.arizona.edu}

\and

\IEEEauthorblockN{Mohammadreza Ebrahimi*}
\IEEEauthorblockA{\textit{School of Information Systems and Management} \\
\textit{University of South Florida}\\
Tampa, USA \\
ebrahimim@usf.edu}

\linebreakand

\IEEEauthorblockN{Hsinchun Chen}
\IEEEauthorblockA{\textit{Department of Management Information Systems} \\
\textit{University of Arizona}\\
Tucson, USA \\
hsinchun@arizona.edu}

}

\maketitle
\thispagestyle{plain}
\pagestyle{plain}

\begin{abstract}
Deep Learning (DL)-based malware detectors are increasingly adopted for early detection of malicious behavior in cybersecurity. However, their sensitivity to adversarial malware variants has raised immense security concerns. Generating such adversarial variants by the defender is crucial to improving the resistance of DL-based malware detectors against them. This necessity has given rise to an emerging stream of machine learning research, Adversarial Malware example Generation (AMG), which aims to generate evasive adversarial malware variants that preserve the malicious functionality of a given malware. Within AMG research, black-box method has gained more attention than white-box methods. However, most black-box AMG methods require numerous interactions with the malware detectors to generate adversarial malware examples. Given that most malware detectors enforce a query limit, this could result in generating non-realistic adversarial examples that are likely to be detected in practice due to lack of stealth. In this study, we show that a novel DL-based causal language model enables single-shot evasion (i.e., with only one query to malware detector) by treating the content of the malware executable as a byte sequence and training a Generative Pre-Trained Transformer (GPT). Our proposed method, MalGPT, significantly outperformed the leading benchmark methods on a real-world malware dataset obtained from VirusTotal, achieving over 24.51\% evasion rate. MalGPT enables cybersecurity researchers to develop advanced defense capabilities by emulating large-scale realistic AMG.

\end{abstract}

\begin{table*} [b!]
\centering
\setlength{\abovecaptionskip}{0pt}
\setlength{\belowcaptionskip}{-10mm}
\begin{center}
\vspace{-4mm}
\caption{Selected Major Append-based AMG Studies}
\begin{tabular}{
|c<{\centering}
|c<{\centering}
|c<{\centering}
|c<{\centering}
|c<{\centering}|}

\hline

\textbf{Year}&\textbf{Author(s)}&\textbf{Data Source}&\textbf{Attack Method}&\textbf{\# of Queries per Malware File}\\

\hline

2021 & Demetrio et al.\cite{demetrio2021functionality} & VirusTotal & Genetic Programming & Unlimited\\

\hline

2020 & Ebrahimi et al.\cite{ebrahimi2020binary} & VirusTotal & Generative RNN & Unlimited\\

\hline

2019 & Castro, Biggio et al.\cite{castro2019armed} & VirusTotal & Gradient descent attack & Unlimited\\

\hline

2019 & Castro, Schmitt et al.\cite{castro2019poster} & VirusTotal & Random perturbations & Unlimited\\

\hline

2019 &  B. et al.\cite{chen2019adversarial} & VirusShare, Malwarebenchmark & Enhanced random perturbations & Multiple\\

\hline

2019 & Dey S. et al.\cite{dey2019evadepdf} & Contagio PDF malware dump & Genetic programming & Multiple\\

%

\hline

2019 & Park et al.\cite{park2019generation} & Malmig \& MMBig & Dynamic Programming & Multiple\\

\hline

2019 & Rosenberg et al.\cite{rosenberg2019defense} & VirusTotal & GAN & Multiple\\

\hline

2019 & Suciu et al.\cite{suciu2019exploring} & VirusTotal, Reversing Labs, FireEye & Append Attack & Single\\

\hline

2018 & Anderson et al.\cite{anderson2018learning} & VirusTotal & Deep RL & Multiple\\

\hline

2018 & Hu \& Tan\cite{hu2018black} & Malwr dataset & Generative RNN & Multiple\\

\hline

\end{tabular}
\vspace{2mm}

{\centering \textbf{Note:} RNN: Recurrent Neural Network; NN: Neural Network; GAN: Generative Adversarial Network; RL: Reinforcement Learning\par} 

\label{lit_overview}
\end{center}
\vspace{-7mm}
\end{table*}

\begin{IEEEkeywords}
Adversarial malware variants, single-shot black-box evasion, deep learning-based language models, generative pre-trained transformers 
\end{IEEEkeywords}
\section{Introduction}

DL-based malware detectors have recently gained attention in the field of cybersecurity due to their ability to identify unseen malware variants without manual feature engineering and expensive dynamic analysis of the behavior of malware instances in a sandbox \cite{raff2018malware}. However, DL-based malware detectors have been shown to be vulnerable to small perturbations in their input, known as adversarial examples \cite{demetrio2019explaining}. The automated generation of such inputs is known as Adversarial Malware example Generation (AMG), which aims to generate functionality-preserving malware variants that mislead these malware detectors. Emulating AMG attacks against malware detectors can help strengthen their malware detection performance \cite{goodfellow2018making}. 

AMG methods can generally be classified into white-box and black-box settings \cite{qiu2019review}. Many AMG methods fall under the white-box classification, where the attackers know the model parameters and architecture of the targeted malware detector. Contrarily, black-box methods assume no attacker knowledge of the model parameters and architecture of the targeted malware detector. Since practical AMG scenarios aligns more with the black-box setting, black-box methods have drawn more attention. 

Most widely-used black-box AMG techniques rely on emulating append attack, an additive approach that injects bytes at non-executable locations in the malware binary. The popularity of append attacks is due to the fact that they are less likely to affect the malware functionality \cite{suciu2019exploring}. To generate adversarial malware variants, these methods require detector feedback, often a Boolean value indicating whether the variant has evaded the malware detector or not. Specifically, current append-based AMG methods require a considerable amount of detector feedback to operate effectively \cite{kolosnjaji2018adversarial} \cite{suciu2019exploring}. Given that real-life malware detectors enforce a query limit, these AMG methods are rendered ineffective due to their query inefficiency. 

The evasion of detectors via only one query, known as \textit{single-shot evasion}, has been well studied in image applications \cite{ilyas2018black}. However, it is understudied in the AMG context. Because the goal of AMG research is to emulate real attacks and improve the performance of malware detectors, there is a vital need for single-shot AMG methods that can emulate realistic adversarial attacks. We expect that well-designed DL-based AMG methods can result in such performance by automatically extracting salient features from the malware sample \cite{awad2018modeling}. Recently, DL-based language models have been shown to effectively extract salient features from sequential data \cite{belletti2019quantifying}. To this end, by viewing a malware executable as a byte sequence, DL-based language models can be utilized to generate evasive malware content \cite{awad2018modeling}. Using DL-based language models' ability to effectively extract salient features for AMG, we seek to increase the likelihood of single-shot evasion using DL-based language models. 

Nevertheless, traditional language models are inefficient at byte sequence generation due to long-range dependencies present in lengthy byte sequences \cite{raff2018malware}. To address this, recent Natural Language Processing (NLP) research has shown DL-based causal language models (CLM) as a viable solution \cite{hosseini2020simple}. Specifically, a recent Causal Deep Language model, Generative Pre-trained Transformer (GPT), has yielded state-of-the-art performance in processing long sequences and high-quality text generation \cite{hosseini2020simple}. Inspired by GPT’s success, we propose a novel AMG framework using a GPT-based language model learned from raw malware content to conduct AMG under a query-efficient threat model that increases the chance of single-shot evasion.

The rest of this paper is organized as follows. First, we review AMG, CLMs, and GPT. Subsequently, we detail the components of our proposed framework. Lastly, we compare the performance of our proposed method with other state-of-the-art AMG methods and highlight promising future directions. 

\section{Literature Review}
Three areas of research are examined. First, we review extant AMG studies as the overarching area for our study. Second, we examine CLM as an effective language model that can facilitate learning patterns in long byte sequences from malware executables. Third, we review GPT as a state-of-the-art causal language model.

\subsection{Adversarial Malware Generation (AMG)}

AMG aims to perturb malware samples and generate variants that evade malware detectors. Among the prevailing AMG methods, append attacks are the most practical due to their high chance of preserving the functionality of the original malware executable \cite{suciu2019exploring}. We  summarize  selected  significant append-based prior work based on their data source, attack method used, and presence of a query limit in Table \ref{lit_overview}.

Three major observation are made from Table \ref{lit_overview}. First, the majority of studies use VirusTotal, an open-source online malware database, as a source of their malware samples \cite{demetrio2021functionality}\cite{castro2019armed}\cite{castro2019poster}\cite{chen2019adversarial}\cite{rosenberg2019defense}\cite{suciu2019exploring}\cite{anderson2018learning}. Second, regarding selected attack methods, a few notable attack methods include simple append attack \cite{suciu2019exploring}, attacking using randomly generated perturbation \cite{castro2019poster}, and attacking using specific perturbations that lowers a malware detector's score \cite{chen2019adversarial}. More advanced methods incorporate machine learning techniques (Genetic Programming \cite{demetrio2021functionality} \cite{dey2019evadepdf}, Gradient Descent \cite{castro2019armed}, and Dynamic Programming \cite{park2019generation}) and implement advanced DL-based techniques (Generative Adversarial Networks \cite{rosenberg2019defense}, Deep Reinforcement Learning \cite{anderson2018learning}, and Generative Recurrent Neural Networks \cite{ebrahimi2020binary}\cite{hu2018black}). Third, and most importantly, while a sizable amount of AMG research either do not limit the number of queries to the malware detector or allow conducting multiple queries, few studies (Suciu et al. \cite{suciu2019exploring}) operate in a single-shot AMG evasion setting. However, the proposed method in \cite{suciu2019exploring} does not use potentially more effective machine learning approaches. Furthermore, more advanced attack methods, such as DL-based ones, require multiple queries per malware file to evade. This is due to the fact that, if not properly designed, DL-based methods require multiple interactions with the detector to receive feedback and learn generating evasive samples through back-propagation. Thus, they are less likely to perform single-shot AMG evasion. 

Overall, we observe that most AMG studies either do not limit the number of queries to the detector or they require multiple queries per malware sample to evade the malware detector. This highlights the inefficiency of their attack methods at extracting salient features and generating evasive samples. 

\subsection{Causal Language Models (CLMs)}
DL-based language models have been shown to effectively extract salient features from sequential data \cite{belletti2019quantifying}. Recently, DL-based language models have been successfully utilized in malware analysis \cite{ebrahimi2020binary}\cite{hu2018black}. Recent AMG research demonstrated the viability of treating malware binaries as a language and generating byte sequences, allowing for automatic perturbation generation in the AMG context \cite{ebrahimi2020binary}. However, due to long-range dependencies in the byte sequences from malware executables, traditional language models become ineffective at learning the patterns present in malware binaries \cite{raff2018malware}. Current NLP research has shown CLM as a promising alternative that can learn patterns in long sequential data. CLMs are characterized by using outputs from previous time steps as inputs in future time steps as it generates bytes. This allows CLMs to reference past information when generating current sequences. Compared to other state-of-the-art language models like BERT \cite{devlin2018bert} and XLNet \cite{yang2019xlnet}, CLMs tend to be less computationally intensive. This allows them to process larger input sequences, making them more suitable for long-range language generation in an AMG context.

\subsection{Generative Pre-trained Transformer}

While CLMs have proved promising in processing sequential data, recent studies have further improved their performance. Specifically, GPT, a recent CLM, has yielded breakthrough performances on NLP tasks \cite{raff2018malware}. GPT is composed of 12 interconnected decoder blocks, each consisting of a self attention layer and a feed-forward neural network as shown in Figure \ref{GPT_Figure}. 

\begin{figure}[h]
\centering
\vspace{-3mm}
\includegraphics[width=0.25\textwidth]{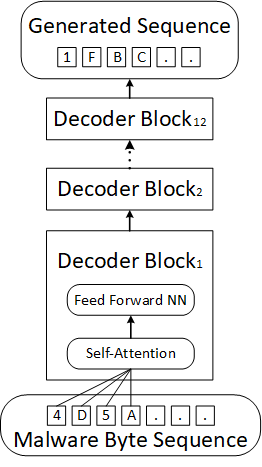}
\vspace{-3mm}
\caption{GPT Architecture Utilized in Byte Generation}
\vspace{-3mm}
\label{GPT_Figure}
\end{figure}

The first layer in each decoder block, known as self attention layer, generates a representation that determines each token's importance in relation to the current token \cite{vaswani2017attention}. From self attention layer, a feed-forward neural network acts as a gateway to pass the obtained representation to the next decoder block \cite{radford2019language}. The same structure holds for each of the 12 decoder blocks, where the last generated sequence is the final output of the $12^{th}$ decoder block. 

GPT’s causal nature and its built-in self-attention mechanism allows it to better learn longer-range patterns when compared to traditional language models \cite{chen2020generative}.

\section{Research Gaps and Questions}

Based on our literature review, two research gaps are identified. First, within the AMG domain, most methods cannot evade malware detectors in a single-shot fashion, causing them to be query inefficient. Second, regarding the methodology, while GPT has shown promising performance in NLP tasks, it is unclear how it could be applied in malware analysis and specifically AMG context.

To address the identified gaps, the following research question is posed: \begin{itemize}
    \item{How can an adversarial causal language model be developed to evade malware detectors with minimal number of queries to maximize the chance of single-shot evasion?} 
\end{itemize}

Motivated by this question, we propose MalGPT, a novel framework to automatically construct adversaries for evading malware detectors in one query utilizing the causal language model GPT2 (a publicly available implementation of GPT).

\section{Research Design}

Following previous AMG studies, we first introduce the threat model under which our proposed MalGPT operates \cite{biggio2013security}. Then, we examine the architecture of MalGPT and its training process. Finally, we introduce our testbed and the targeted malware detector used in MalGPT's training and evaluation. 

\subsection{Threat Model}

A threat model is a systematic representation of cyber attacks \cite{carlini2019evaluating}\cite{biggio2013security}. Since the goal of our study is evading DL-based detectors with only one query and without accessing the internal model parameters of the malware detector, our threat model focuses on a single-shot, black-box setting. Accordingly, three major components of our threat model are:

\begin{itemize}
  \item \textbf{Adversary’s Goal:} Evade DL-based malware detectors in a single shot. That is, the evasive adversarial malware variants that are generated after one interaction with the detector do not count towards model's success.
  \item \textbf{Adversary’s Knowledge:}
  \begin{itemize}
      \item Structure and parameters of malware detector model are unknown to the attacker.
      \item Attacker does not have access to the confidence score produced by malware detector (fully black-box attack).
  \end{itemize}
  \item \textbf{Adversary's Capability:} The adversary applies functionality-preserving append modifications on malware binaries.
  \begin{itemize}
      \item Consistent with past AMG studies \cite{ebrahimi2020binary} \cite{kolosnjaji2018adversarial}, the size of the modifications must stay under 10 KB to maintain the stealth of the generated malware variant.
  \end{itemize}
\end{itemize}

\begin{figure*}[t]
    \centering
    \vspace{-3mm}
    \includegraphics[width=0.85\textwidth]{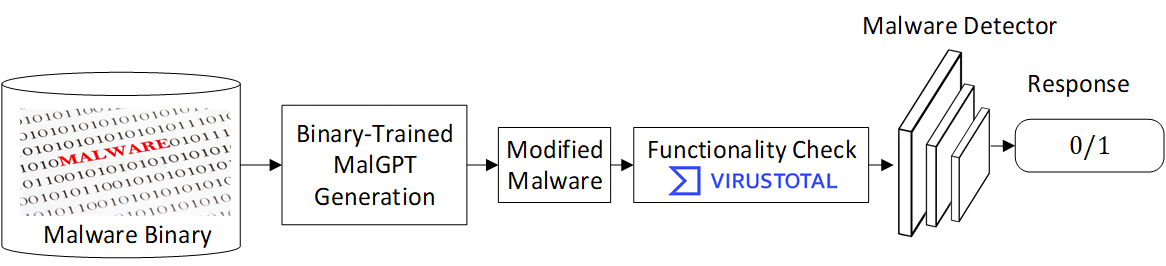}
    \vspace{-3mm}
    \caption{Abstract View of MalGPT Malware Evasion Framework}
    \vspace{-3mm}
    \label{Framework_Architecture}
\end{figure*}

\subsection{MalGPT Architecture}\label{ModelFrameworkSection}

To realize this threat model, MalGPT employs a GPT2 language model that is trained to generate benign-looking byte sequences. The trained language model is utilized to generate evasive and functionality-preserving variants of existing known malware executables as depicted in Figure \ref{Framework_Architecture}. This process is detailed in 5 steps: 

\begin{itemize}
    \item \textit{Step 1:} The binary content of a malware sample is fed into the trained GPT2 model.
    \item \textit{Step 2:} The model generates a file-specific byte sequence. 
    \item \textit{Step 3:} The generated sequence is added to the original malware sample, resulting in a new malware variant.
    \item \textit{Step 4:} The new malware variant is examined for functionality using the VirusTotal API. 
    \item \textit{Step 5:} After confirming its functionality, the generated variant attempts to evade a malware detector in a single query.
\end{itemize}

\subsection{MalGPT Model Training}\label{ModelTrainingSection}

In order to generate benign-looking sequences, MalGPT's GPT2 model is trained on a set of benign files. Figure \ref{Model_Training} provides an illustration of this process, with the final trained model being incorporated into Figure \ref{Framework_Architecture} as the Binary-Trained MalGPT Generation. 

\begin{figure}[!h]
    \centering
    \includegraphics[width=0.5\textwidth]{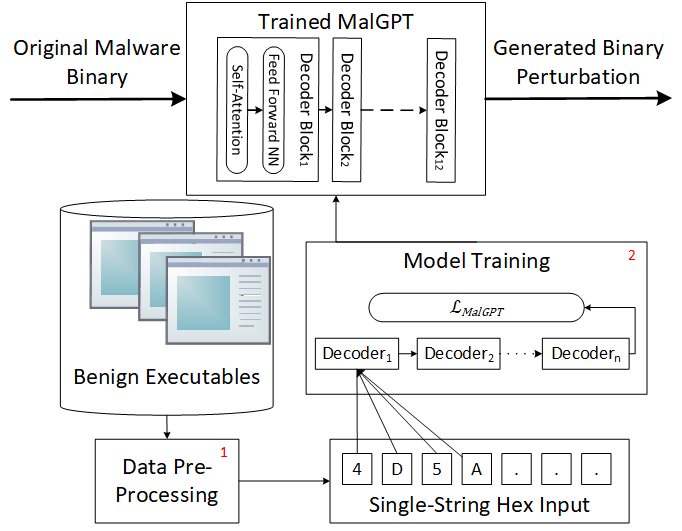}
    \caption{Illustration of MalGPT's Model Training Process}
    \label{Model_Training}
\end{figure}

The model is trained in a two-step process. In Step 1, benign files are first converted to a single hex string delimited by sets of four characters (e.g., `AA04 FF44 ...') and fed into the GPT2 model. Step 2 has the model automatically extracting salient features from the hex string and learning how to generate benign-looking byte sequences. After repeating Step 2 for 1,000 training iterations, the process results in a trained model that takes malware byte sequences as input to generate benign-looking, file-specific perturbations.

\subsection{Testbed and Targeted Malware Detector}
As stated in prior sections, MalGPT requires both benign and malicious files to be trained and to evade malware detectors. To this end, following the approach in \cite{raff2018malware}, we collected 13,554 benign Microsoft Windows system files for MalGPT to learn from. Additionally, we obtained 6,307 malicious samples from VirusTotal in eight major malware categories. Table \ref{testbed_table} summarizes the distribution of these categories along with their description and examples.

\begin{table}[!ht]
\centering
\begin{center}
\vspace{-4mm}
\caption{Malware Samples in our Testbed}
\vspace{-2mm}
\begin{tabular}{
|m{1.2cm}<{\centering}
|m{3.0cm}<{\centering}
|m{1.5cm}<{\centering}
|m{0.6cm}<{\centering}
|}
\hline
\textbf{Malware Category} &\textbf{Description} &\textbf{Examples} &\textbf{\# of Files}\\

\hline
\textbf{Adware} & Shows unwanted ads and force internet traffic to sites & eldorado, razy, gator & 1,947\\

\hline
\textbf{Backdoor} & Negates normal authentications to access the host & lunam, rahack, symmi & 678\\

\hline
\textbf{Botnet} & A network of bots connected through the internet & virut, salicode, sality & 526\\

\hline
\textbf{Dropper} & Secretly installs other malwares on the host & dinwod, gepys, doboc & 904\\

\hline
\textbf{\makecell{Ransom-\\ware}} & Encrypts data and files, restricting access and usage until decrypted by malware authors & vtflooder, msil, bitman & 900\\

\hline
\textbf{Rootkit} & Grants admin privilege to malware author & onjar, dqqd, shipup & 53\\

\hline
\textbf{Spyware} & Allows malware authors to steal personal information covertly & mikey, qqpass, scar & 640\\

\hline
\textbf{Virus} & Corrupts files on the host system & nimda, shodi, hematite & 659\\
\hline
\textbf{Total} & - & - & \textbf{6,037}\\
\hline

\end{tabular}
\label{testbed_table}
\vspace{-4mm}
\end{center}
\end{table}

\section{Evaluation}

\subsection{Experiment Design}
Through consulting with two malware analysis experts as well as the popularity in malware analysis community, we selected MalConv as one of the most highly reputable DL-based malware detectors \cite{raff2018malware}. To evaluate performance, consistent with \cite{fleshman2018non}, we adopted evasion rate as the most common performance metric in AMG research. The evasion rate is defined via the following equation \cite{ebrahimi2020binary}:

\begin{equation}
  Evasion\ Rate = \frac{|E\cap F|}{N}
\end{equation}
where $E$ and $F$ denote the sets of evasive and functional modified malware samples generated from the AMG method, respectively. $N$ represents the total number of malware samples given as input to the AMG method. To evaluate MalGPT's performance, we conducted several benchmark experiments under the constraints of our threat model (i.e., single-shot evasion, black-box setting, and 10 KB maximum 
append size). Table \ref{benchmark_table} presents the description for each selected benchmark method. 

\begin{table}[!ht]
\centering
\vspace{-4mm}
\begin{center}
\caption{Overview of Benchmark Experiment Methods}
\vspace{-2mm}
\begin{tabular}{
|m{1.5cm}<{\centering}
|m{3cm}<{\centering}
|m{3cm}<{\centering}|}
\hline
\textbf{Method} & \textbf{Description} & \textbf{Reference(s)}\\

\hline
\textbf{Random Append} & Randomly appends bytes to malware sample. & Suciu et al., 2019 \cite{suciu2019exploring}; Castro, Schmitt et al., 2019 \cite{castro2019armed}\\

\hline
\textbf{Benign Append} & Appends sections of bytes from benign files to malware sample. & Castro, Biggio et al., 2019 \cite{castro2019poster}\\

\hline
\textbf{Enhanced Benign Append} & Appends bytes that lower the confidence score the most. & Chen B. et al., 2019 \cite{chen2019adversarial}\\

\hline
\textbf{MalRNN} & Appends a byte sequence generated by an RNN model trained on benign files & Ebrahimi et al., 2020 \cite{ebrahimi2020binary}\\

\hline

\end{tabular}
\label{benchmark_table}
\vspace{-4mm}
\end{center}
\end{table}

Each AMG method was performed on the eight individual malware categories in the testbed as well as the entire testbed (i.e., all 6,037 malware executables) to gauge its efficacy at single-shot AMG evasion based on evasion rate.

\subsection{Experiment Results}
Table \ref{ex_result} shows MalGPT’s performance compared to state-of-the-art AMG benchmarks against MalConv under the constraints of the defined threat model. The row denoted by `Total' corresponds to the performance on the entire testbed.

\begin{table}[!ht] 
\centering
\setlength{\abovecaptionskip}{0pt}
\setlength{\belowcaptionskip}{-10mm}
\begin{center}
\vspace{-4mm}
\caption{Experiment Results}
\vspace{-2mm}
\begin{tabular}{
|m{1cm}<{\centering}
|m{1cm}<{\centering}
|m{1cm}<{\centering}
|m{1.25cm}<{\centering}
|m{1cm}<{\centering}
|m{1cm}<{\centering}|}

\hline
\textbf{Category} & \textbf{Random Append} & \textbf{Benign Append} & \textbf{Enhanced Benign Append} & \textbf{MalRNN} & \textbf{MalGPT}\\

\hline
\textbf{Adware} & 2\% & 0.87\% & 15.51\% & 4.16\% & \textbf{25.89\%*}\\

\hline
\textbf{Backdoor} & 2.06\% & 0.74\% & 21.98\% & 0.44\% & \textbf{18.86\%}\\

\hline
\textbf{Botnet} & 2.47\% & 1.14\% & 21.86\% & 6.08\% & \textbf{25.86\%*}\\

\hline
\textbf{Dropper} & 3.32\% & 2.32\% & 16.48\% & 4.2\% & \textbf{27.43\%*}\\

\hline
\textbf{\makecell{Ransom-\\ware}} & 3.78\% & 0.11\% & 14.44\% & 0.89\% & \textbf{20.33\%*}\\

\hline
\textbf{Rootkit} & 1.89\% & 3.77\% & 3.77\% & 5.66\% & \textbf{24.53\%*}\\

\hline
\textbf{Spyware} & 2.5\% & 1.88\% & 11.25\% & 4.38\% & \textbf{22.97\%*}\\

\hline
\textbf{Virus} & 4.4\% & 2.43\% & 12.29\% & 10.17\% & \textbf{28.38\%*}\\

\hline
\textbf{Total} & 2.79\% & 1.27\% & 15.86\% & 4.12\% & \textbf{24.51\%*}\\

\hline

\end{tabular}
\vspace{1mm}

\centering \textbf{Note:} P-Values are significant at 0.05.

\label{ex_result}
\vspace{-5mm}
\end{center}
\end{table}

The asterisks in Table \ref{ex_result} denote the statistical significance obtained from paired $t$-test at P-value equal to or less than 0.05 between the results of MalGPT and the second-best performing benchmark method in each respective category. The results show approximately a 20\% performance improvement over the recently proposed state-of-the-art malware language model, MalRNN (4.12\% vs. 24.51\%). Moreover, MalGPT shows an approximately 7\% performance improvement at single-shot evasion over the second-best method, Enhanced Benign Append. Overall, from Table \ref{ex_result} it is shown that MalGPT attains the best performance on the entire dataset (24.51\%) and outperforms other benchmark methods in almost all categories (except backdoor). The significantly high performance of MalGPT compared to the benchmark methods suggests that, as expected, the high-quality representations obtained by the GPT2 component in our model effectively increase the chance of single-shot evasion.  

In addition to comparison with other AMG benchmark methods, it is useful to compare MalGPT's performance across all eight malware categories. Figure \ref{ExpResults} depicts the evasion rate of MalGPT for each malware category a long with the evasion rate across the entire 6,307 malware executables (denoted by `Total').

\begin{figure}[!h]
    \centering
    \vspace{-3mm}
    \includegraphics[width=0.5\textwidth]{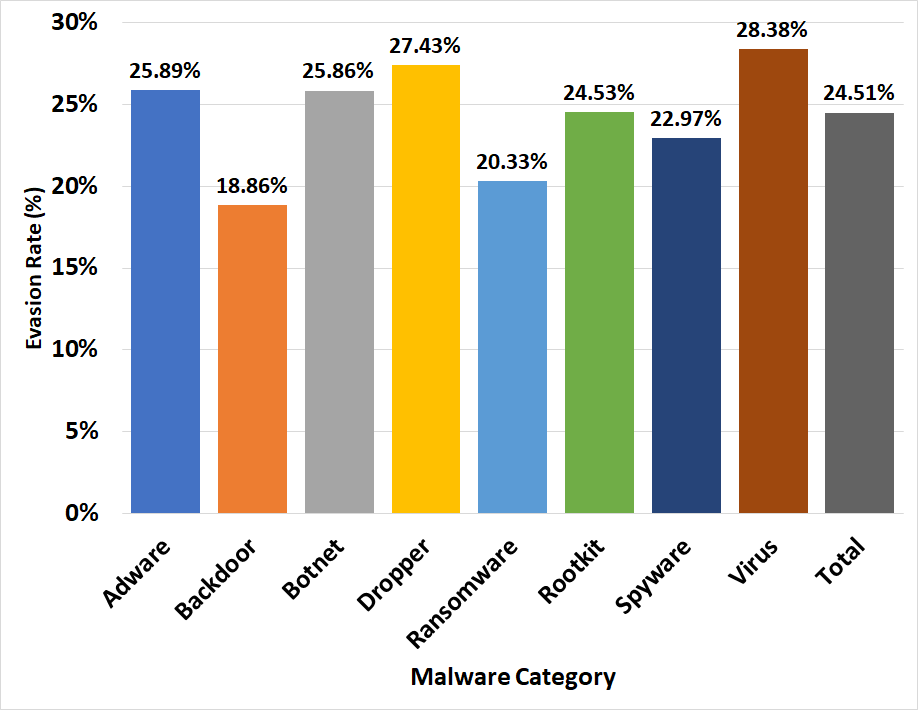}
    \vspace{-3mm}
    \caption{MalGPT's Evasion Rate for each Malware Category and entire dataset}
    \vspace{-3mm}
\label{ExpResults}
\end{figure}

From Figure \ref{ExpResults}, we have a few observations. While MalGPT attains an evasion rate of 24.51\% across all 6,307 malware samples in our testbed, several categories noticeably deviate from this trend. On the upper end, both Dropper and Virus have a high evasion rate at 27.43\% and 28.38\%, respectively. Conversely, both Backdoor and Ransomware have a lower evasion rate at 18.86\% and 20.33\%, respectively. These results suggest that Dropper and Virus are more sensitive to AMG append attacks while Backdoor and Ransomware may be less sensitive to such attacks. One possible explanation could be because of the malware file size. Droppers are usually small files that download other malicious files through a link after gaining access to a host machine. Likewise, Viruses are often small scripts that seek to corrupt a host machine. As such, both categories feature smaller file size allowing a 10 KB perturbation generated by MalGPT to have a larger effect and thus a higher evasion chance. Conversely, both Backdoor and Ransomware often need large, complex programs (e.g., encryption procedures) to achieve their malicious goals. As such, MalGPT's 10 KB perturbation could be less effective in larger files, thus making evasion more difficult. This aligns with the intuition that DL-based malware detectors are more likely to be evaded with proportionally larger AMG perturbations with respect to the original file size. 


Overall, the experiment results suggest that our proposed approach of implementing GPT into an AMG framework significantly improves the chance of single-shot evasion. Additionally, our results show the deficiency of current AMG methods to operate effectively in a single-shot threat model. This highlights their excessive reliance on querying a malware detector multiple times, which renders them ineffective in practice when realistic restrictions are applied to the number of allowed queries. 

\section{Conclusion and Future Directions}

AMG research has gained popularity as a way to better understand and combat malware attacks. However, current AMG methods are rendered ineffective in real-world settings due to their reliance on multiple malware detector queries and the frequent implementation of query limits on malware detectors in practice. Leveraging GPT, we propose a novel framework for evading DL-based malware detectors that operationalizes a single-shot black-box evasion threat model. The proposed MalGPT framework utilizes GPT's ability to extract salient features from long-range dependencies in byte sequences extracted from malware executable content and generate benign-looking byte sequences for single-shot AMG evasion.
Our MalGPT was evaluated on eight major malware categories. MalGPT significantly outperformed all benchmark methods, demonstrating its ability to operate effectively in a single-shot setting where other AMG methods cannot.

Our proposed research could be further extended by incorporating other views of sequential malware data such as malware source code (in addition to raw binary content). Multi-view deep learning methods that can utilize information from both executable's raw content and source code are anticipated to yield better evasion performance.
\bibliographystyle{IEEEtran}
\bibliography{main}

\vspace{12pt}
\color{red}

\end{document}